\documentstyle[preprint,cite,epsf,aps]{revtex}
\begin{document}

\title{\Large  The Three-Dimensional Noncommutative Nonlinear Sigma  Model
in Superspace}

\author{H. O. Girotti$^{\,a}$, M. Gomes$^{\,b}$, A. Yu. Petrov$^{b,c}$,
  V. O. Rivelles$^{\,b}$ and A. J. da Silva$^{\,b}$}  
\address{$^{a\,}$Instituto de F\'\i sica, Universidade Federal do Rio Grande
do Sul\\ Caixa Postal 15051, 91501-970 - Porto Alegre, RS, Brazil\\ E-mail:
hgirotti@if.ufrgs.br}
\address{$^{b\,}$Instituto de F\'\i sica, Universidade de S\~ao Paulo\\
 Caixa Postal 66318, 05315-970, S\~ao Paulo - SP, Brazil\\
E-mail: mgomes, petrov, rivelles, ajsilva@fma.if.usp.br}
\address{$^{c}$Department of Theoretical Physics,\\
Tomsk State Pedagogical University\\
Tomsk 634041, Russia\\
E-mail: petrov@tspu.edu.ru}

\maketitle

\begin{abstract}
We study the superspace formulation of the noncommutative nonlinear
supersymmetric $O(N)$ invariant sigma-model in 2+1 dimensions.  We
prove that the model is renormalizable to all orders of 1/N and
explicitly verify that the model is asymptotically free.
\end{abstract}

\vspace*{2.0cm}

Noncommutative field theories present many unusual
properties. As a consequence of the noncommutativity,
high momentum modes do not decouple from the physics at large
distances leading to the appearance of infrared poles even in theories
without massless particles. Being nonintegrable, these infrared
singularities destroy the usual perturbative expansions. This fact
has motivated many studies on the renormalization
properties of noncommutative field theories \cite{Minwalla,Matusis,Hayakawa,Arefeva2,Arefeva,Martin,Grosse,Armoni,Bonora,Sheikh,Chepelev,Gracia}.

Based on previous experience with commutative field theories one may
wonder if supersymmetry is able to solve this problem without
jeopardizing unitarity. This has been proven to be correct for the
Wess--Zumino model and also, at least to the subleading order of
$1/N$, for the nonlinear sigma model \cite{Girotti,prev}. In this
context, the use of the superspace formalism \cite{Popp} has shown to
be a very powerful tool to investigate
noncommutative supersymmetric theories \cite{Zanon,ours}.  In the
present work, using superfield techniques to accommodate the
intricacies of the Moyal product, we will demonstrate that the
noncommutative nonlinear sigma model is renormalizable to all orders
of $1/N$.  Furthermore, the renormalization group equations are
analyzed and we prove that the model is asymptotically free.
 
The action of the  noncommutative $O(N)$ sigma model in three-dimensional
space-time is \cite{prev,Koures}

\begin{equation}
\label{1}
S=\int d^3 x d^2\theta \frac{1}{2}\left[\Phi_j D^2 \Phi_j-\Sigma
\star(\Phi_j\star\Phi_j-\frac{N}{g})\right],
\end{equation}

\noindent
where $\Phi_i$, $i=1, \ldots N$, are real superfields and $\Sigma$ is a
Lagrange multiplier  superfield which enforces the constraint
$\Phi_j\Phi_j=\frac{N}{g}$. Just for
reference  and to make contact with our previous work\cite{prev}, we quote
the field component expansions for $\Phi_j$ and $\Sigma$:

\begin{eqnarray}
\Phi_j&=& \varphi_j + \bar \theta \psi_j+ \frac12 \bar \theta\theta F_j,\\
\Sigma &=& \sigma + \bar \theta \xi + \frac12 \bar \theta \theta\lambda,
\end{eqnarray}

\noindent
where $\psi$ is a $N$ component Majorana spinor and $\varphi$ and $F$
are $N$ components scalar fields.

We are interested in a massive phase where $\Sigma$ acquires a
nonvanishing vacuum expectation value. Thus, replacing $\Sigma \rightarrow
\Sigma +m$, where $m$ is the vacuum expectation value of the original
$\Sigma$, we obtain the new action 

\begin{equation}
\label{2}
S=\int d^3 x d^2\theta \frac{1}{2}\left [\Phi_j (D^2 -m)\Phi_j
-\Sigma\star(\Phi_j\star\Phi_j-\frac{N}{g})\right].
\end{equation}

\noindent
Using this expression it is straightforward to verify that the propagator
for the basic superfields is

\begin{equation}\label{3}
<\tilde \Phi_i(p_1,\theta_1)\tilde\Phi_j(p_2,\theta_2)>=i\frac{D^2+m}{p_{1}^{2}-m^2}\delta_{ij}\delta^3(p_1-p_2)\bar\delta_{12},
\end{equation}

\noindent
where we have introduced the notation $\bar\delta_{12}=
\delta(\bar\theta_1 - \bar \theta_2)\delta(\theta_1-\theta_2)$, $D^2=
\frac12 \bar D D$, the supercovariant derivative $D$ is

\begin{equation}\label{4}
D= \frac{\partial\phantom a}{\partial \theta} - i \overline \theta \not \!
\partial,
\end{equation}

\noindent
and $\bar D_\alpha =\gamma^{0}_{\alpha\beta} D_\beta$. The interaction
vertex is given by

\begin{eqnarray}
\label{star}
\int d^5 z \Sigma \star\Phi_j\star\Phi_j&=&\int d^2\theta \int
\frac{d^3k_1 d^3k_2 d^3k_3}{(2\pi)^9}
\exp (-i \sum_{a<b} k_a\wedge
k_b) \tilde{\Sigma} (k_3,\theta)\tilde{\Phi}_j (k_1,\theta) 
\tilde{\Phi}_j (k_2,\theta)\nonumber\\&\times&
(2\pi)^3 \delta(k_1+k_2+k_3)\equiv\nonumber\\
&\equiv&\int d^2\theta\int\frac{d^3k_1 d^3k_2}{(2\pi)^6}	
\cos(k_1\wedge
k_2)\tilde{\Sigma}(-k_1-k_2,\theta)\tilde{\Phi}_i(k_1,\theta)\tilde{\Phi}_i
(k_2,\theta).
\end{eqnarray}

\noindent
Here, $k_a\wedge k_b=\frac{1}{2}k_{a\mu}\Theta^{\mu\nu}k_{b\nu}$,
$\Theta_{\mu\nu}$ is the noncommutativity matrix and $\tilde{\Phi_i}$
and $\tilde\Sigma$ denote the Fourier transforms of the superfields
$\Phi_i$ an $\Sigma$, respectively.

To find the effective propagator for the auxiliary field, let us
consider the supergraph shown in Fig. 1. It is straightforward to verify
that it contributes 

\begin{equation}\label{5}
{N}\int d^2\theta \tilde \Sigma (k,\theta) (D^2+2m)\tilde \Sigma (-k,\theta) 
I^{-1}(k^2),
\end{equation}

\noindent
where

\begin{equation}
I^{-1}(k^2)=\frac12\int \frac{d^3 l}{(2\pi)^3}
\frac{\cos^2 (k\wedge l)}{(l^2-m^2)((k+l)^2-m^2)}.
\end{equation}

\noindent
Using the relation \cite{Gelfand}

\begin{equation}
\label{macd}
\int \frac{d^n k}{(2\pi)^n}\frac{e^{ik\wedge p}}{(k^2+M^2)^\lambda}=
\frac{M^{n/2-\lambda}}{2^{\lambda-1}(2\pi)^{n/2}\Gamma[\lambda]}
\frac{K_{n/2-\lambda}(\sqrt{-M^2 p\circ p})}{{(\sqrt{- p\circ p})}^{n/2
-\lambda}},
\end{equation}

\noindent
where $p\circ p=p_{\mu}(\Theta^2)^{\mu\nu}p_{\nu}$, we get

\begin{eqnarray}
\label{i}
I^{-1}(k^2)&=&\frac{i}{32\pi}
\int_0^1\frac{dx}{\sqrt{m^2-k^2x(1-x)}}
\Big[1-i\sqrt[4]{4k\circ k (m^2-k^2x(1-x))}\nonumber\\&\times&
\sqrt{\frac2\pi}K_{-1/2}
(\sqrt{4k\circ k (m^2-k^2x(1-x))})
\Big].
\end{eqnarray}

\noindent
For large $k$ the last term either exponentially decreases or strongly
oscillates; in both cases, its presence in a Feynman integral will lead to
a finite result.  For practical purposes, the dominant asymptotic behavior 
can be then taken as

\begin{equation}
\label{asymp1}
I^{-1}(k^2)\simeq\frac{i}{32\pi}\int_0^1\frac{dx}{\sqrt{m^2-k^2x(1-x)}}
\simeq \frac{i}{32\sqrt{-k^2}}.
\end{equation}

\noindent
In the sequel we are going to discuss the ultraviolet behavior of the $1/N$
expansion for the model. To that end we find convenient to replace
$I^{-1}(k^2)$ by the expression
 
\begin{equation}
\label{asymp}
I^{-1}(k^2)\simeq\frac{i}{32\sqrt{-k^2+4m^2}},
\end{equation}

\noindent
which has the same asymptotic behavior at $k\to\infty$
as (\ref{asymp1}). Of course, such replacement does not alter
the leading UV behavior of Feynman integrals. Proceeding in this way,
we obtain

\begin{equation}
<\tilde \Sigma(k_1,\theta_1)\tilde \Sigma(k_2,\theta_2)>=-
\frac{1}{2 N}I(k_{1}^{2})\frac{D^2-2m}{k^{2}_{1}-4m^2}\delta^5_{12}\simeq
-\frac{16i}{N}\frac{D^2-2m}{\sqrt{-k^{2}_{1}+4m^2}}\delta^3(k_1-k_2)
\bar \delta_{12}.
\end{equation}

The superficial degree of divergence associated to a generic Feynman
supergraph $\gamma$ having, respectively, $n_\Phi$ and $n_\Sigma$
internal $\Phi$ and $\Sigma$ lines is given by

\begin{equation}\label{7}
d(\gamma) = 3 L - 2n_\Phi-n_\Sigma + N_{D^2},
\end{equation}

\noindent
where $L$ is the number of loops and $N_{D^2}$ is the maximum number of $D^2$
factors which turn into loop momenta after the $\theta$ integration. Due
to the properties of the $D^2$ derivatives \cite{Grisaru}, this number is

\begin{equation}\label{8}
N_{D^2}=n_\Phi+n_\Sigma -L= V-1,
\end{equation}

\noindent
where $V$ is the number of vertices. Thus,

\begin{equation}
d(\gamma) =2 -N_{\Sigma}-\frac{N_{\Phi}}{2},
\end{equation}

\noindent
where $N_{\Sigma}$ and $N_{\Phi}$ are number of $\Sigma$ and
$\Phi_i$ external lines, respectively. One should stress that (\ref{8})
is actually an upper limit on the number of available $D^2$ factors
since, after taking one $D^2$ factor for each loop, the number
of those that may be converted into momentum must be even.

Right away one sees that there is no quadratic divergences except for
vacuum diagrams and, as we shall see shortly, there are neither linear
divergences; hence this theory is free of the nonintegrable infrared
divergences which spoil the usual perturbative expansions.

Before embarking into the general discussion, we shall analyze all
possible cases of divergences at the leading order of $1/N$. Linear
divergences may arise in graphs having either $N_\Sigma=0$, $N_\Phi=2$
or $N_\Sigma=1$, $N_\Phi=0$.  The last situation corresponds to the
tadpole graph which, as we know, should be absent once the mass $m$
for the $\Phi$ field has been generated. Let us then consider the
other possibility, which corresponds to the leading radiative
correction to the $\Phi$ field propagator depicted in Fig. 2. By
partially integrating, we may transfer the $D^2$ derivative from one
of the propagators to the other lines. However, to get a nonvanishing
result, only one factor of $D^2$ may survive in this process. Then,
one of the $D^2$ has to be transferred to the external line and so the
degree of divergence is lowered by one. At one-loop, just a
logarithmic divergence may remain. Explicitly, the supergraph
corresponds to

\begin{eqnarray}
I_1&=&\frac{16}{N}\int \frac{d^3 p}{(2\pi)^3} d^2\theta_1 d^2\theta_2
\tilde \Phi_i(-p,\theta_1)\tilde \Phi_i(p,\theta_2)
\nonumber\\&\times&
\int \frac{d^3k}{(2\pi)^3}\cos^2(k\wedge p)
\frac{(D^2-2m)\bar\delta_{12}(D^2+m)\bar\delta_{12}}{((k+p)^2-m^2)\sqrt{-k^2+4m^2}},
\end{eqnarray}

\noindent
which, after the aforementioned $D$-algebra transformation, can be casted as

\begin{eqnarray}
I_1&=&\frac{8}{N}\int \frac{d^3 p}{(2\pi)^3} d^2\theta\,
\tilde \Phi_i(-p,\theta)(D^2-m)\tilde\Phi_i(p,\theta)
\nonumber\\&\times&
\Big[\int \frac{d^3 k}{(2\pi)^3}
\frac{1}{(k^2-m^2)\sqrt{-k^2+4m^2}}+fin\Big],
\end{eqnarray}

\noindent
where $fin$ indicates a finite contribution. In this paper, all divergent 
integrals are to be understood as being dimensionally
regularized. 
By integrating over $k$ and 
Fourier transforming with respect to the external momentum, we obtain

\begin{equation}\label{6}
I_1=-\frac{4i}{\pi^{2}N}\int d^5 z \Phi_i (D^2-m)\Phi_i
\left(\frac{1}{\epsilon}+fin\right).
\end{equation}

\noindent  
Notice that in the commutative case the divergence of the $\Phi$
propagator is the same as in (\ref{6}). To finalize this preliminary
discussion, we need to consider the leading corrections to the three
point function. The relevant diagrams are shown in Fig. 3. The
contributions of these supergraphs are, respectively,

\begin{eqnarray}
I_{3a}&=&-
\frac{16}{N}\int \frac{d^3p_1 d^3p_2}{(2\pi)^6}\int
d^2\theta_1 d^2\theta_2 d^2\theta_3 \tilde \Phi_i(p_1,\theta_1)\tilde\Phi_i(-p_2,\theta_2)
\tilde\Sigma(p_2-p_1,\theta_3)\nonumber\\&\times&
(D^2-2m)\bar\delta_{12}(D^2+m)\bar\delta_{13}(D^2+m)\bar\delta_{32}
\nonumber\\&\times&
\int\frac{d^3k}{(2\pi)^3}
\frac{\cos(k\wedge p_1)\cos(k\wedge p_2)\cos((k+p_1)\wedge(p_1-p_2))}
{\sqrt{-k^2+m^2}[(k+p_1)^2-m^2][(k+p_2)^2-m^2]}
\end{eqnarray}

\noindent
and

\begin{eqnarray}
I_{3b}&=&
\frac{(16)^2i}{N}\int \frac{d^3p_1 d^3p_2}{(2\pi)^6}\int
d^2\theta_1 d^2\theta_2 d^2\theta_3 d^2\theta_4 d^2\theta_5\tilde\Phi_i(p_1,
\theta_1)\tilde\Phi_i(-p_2,\theta_2)\tilde\Sigma(p_2-p_1,\theta_5)\nonumber\\&\times&
(D^2+m)\bar\delta_{12}(D^2-2m)\bar\delta_{23}(D^2+m)\bar\delta_{34}
(D^2-2m)\bar\delta_{41}(D^2+m)\bar \delta_{45}(D^2+m)\bar\delta_{53}
\nonumber\\&\times&
\int\frac{d^3k d^3 l}{(2\pi)^6}
\frac{
\cos(k\wedge p_1)\cos(k\wedge p_2)\cos((k+p_1)\wedge l)\cos((k+p_2)\wedge l)
}
{
\sqrt{-(k+p_1)^2+m^2}\sqrt{-(k+p_2)^2+m^2}(k^2-m^2)(l^2-m^2)}
\nonumber\\&\times&\frac{\cos((k-l+p_1)\wedge (p_1-p_2))}
{[(k-l+p_1)^2-m^2][(k-l+p_2)^2-m^2]
}.
\end{eqnarray}

\noindent
Let us first consider $I_{3a}$. It is straightforward to see that
$\cos(k\wedge p_1)\cos(k\wedge p_2)\cos((k+p_1)\wedge (p_1-p_2))=
\frac{1}{4}\cos(p_1\wedge p_2)+\ldots$, where the dots indicate terms that depend on the internal momenta and thus will give finite
contributions. Notice also that after integrating on $\theta_3$

\begin{equation}
(D^2-2m)\bar \delta_{12}(D^2+m)\bar \delta_{13}(D^2+m)\bar\delta_{32}\rightarrow k^2\bar\delta_{12}+\ldots,
\end{equation}

\noindent
due to the properties of the supercovariant derivatives. As a result of these manipulations, we can isolate the divergent part of $I_{3a}$, 

\begin{eqnarray}
I_{3a}&=&
-\frac{4}{N}
\int \frac{d^3p_1 d^3p_2}{(2\pi)^6}\int
d^2\theta \tilde\Phi(p_1,\theta)\tilde\Phi_i(-p_2,\theta)\tilde\Sigma(p_2-p_1,\theta)
\nonumber\\&\times&\cos(p_1\wedge p_2)
\int\frac{d^3k}{(2\pi)^3}
\frac{k^2}
{\sqrt{-k^2+m^2}[(k+p_1)^2-m^2][(k+p_2)^2-m^2]}+\ldots,
\end{eqnarray}

\noindent
so that the pole term is

\begin{eqnarray}
\label{phi}
I_{3a}^{div}&=&\frac{2i}{N\pi^2 \epsilon}
\int \frac{d^3p_1 d^3p_2}{(2\pi)^6}\int
d^2\theta \tilde\Phi_i(p_1,\theta)\tilde\Phi_i(-p_2,\theta)\tilde
\Sigma(p_2-p_1,\theta)
\cos(p_1\wedge
p_2).
\end{eqnarray}

\noindent
To obtain the corresponding divergence in the commutative case one should
multiply this result by two and replace the cosine factor by one. 

Concerning $I_{3b}$, first notice that

\begin{eqnarray}
&&\cos(k\wedge p_1)\cos(k\wedge p_2)\cos((k+p_1)\wedge l)\cos((k+p_2)\wedge l)
\cos((k-l+p_1)\wedge (p_1-p_2))\nonumber \\
&&=\frac{1}{16}\cos(p_1\wedge p_2)+\ldots\, .
\end{eqnarray}

\noindent
After performing the $\theta$ integrals, supercovariant derivative manipulations furnish now

\begin{eqnarray}
&&(D^2+m)\bar\delta_{12}(D^2-2m)\bar\delta_{23}(D^2+m)\bar\delta_{34}
(D^2-2m)\bar\delta_{41}(D^2+m)\bar\delta_{45}(D^2+m)\bar\delta_{53}\nonumber\\
&&\rightarrow (k+p_2)^2(k-l+p_2)^2+\ldots \,.
\end{eqnarray}

\noindent
Using these results, one arrives at

\begin{eqnarray}
I_{3b}&=&
\frac{16i}{N}\int \frac{d^3p_1 d^3p_2}{(2\pi)^6}\int
d^2\theta \tilde\Phi_i(p_1,\theta)\tilde\Phi_i(-p_2,\theta)
\tilde \Sigma(p_2-p_1,\theta)\cos(p_1\wedge p_2)
\nonumber\\&\times&
\int\frac{d^3k d^3 l}{(2\pi)^6}
\frac{(k+p_2)^2(k-l+p_2)^2}
{\sqrt{-(k+p_1)^2+m^2}\sqrt{-(k+p_2)^2+m^2}(k^2+m^2)(l^2-m^2)}
\nonumber\\&\times&
\frac{1}{
[(k-l+p_1)^2-m^2][(k-l+p_2)^2-m^2]
}+fin,
\end{eqnarray}

\noindent
which contains the following divergent part

\begin{eqnarray}
I_{3b}^{div}&=&
\frac{i}{\pi^2N\epsilon}\int \frac{d^3p_1 d^3p_2}{(2\pi)^6}\int
d^2\theta \tilde\Phi_i(p_1,\theta)\tilde\Phi_i(-p_2,\theta)
\tilde\Sigma(p_2-p_1,\theta)\cos(p_1\wedge p_2).
\end{eqnarray}

\noindent
Summarizing, the divergent part of the vertex correction is

\begin{equation}
\label{sigma}\label{9}
S_{\Sigma}=\frac{3i}{N\pi^2\epsilon}\int d^5 z \Sigma\star\Phi_i\star\Phi_i. 
\end{equation}

\noindent
In the commutative case the corresponding result is  

\begin{equation}
\frac{8i}{N\pi^2\epsilon}\int d^5 z \Sigma\Phi_i\Phi_i, 
\end{equation}

\noindent
so that, in view of (\ref{6}), it may be eliminated by just a wave
function renormalization of the $\Phi$ field. Unlike the commutative
case, however, the renormalization of the vertex  requires here
also a wave function renormalization for the auxiliary
field $\Sigma$.  From a formal viewpoint, this is caused by the
presence of additional factors $1/2^{V-1}$ in the planar
contributions. These modifications are, of course, a consequence of the
specific nonlocality induced by the Moyal products.

To complete the discussion of the renormalization at leading order we
should examine the possible divergences associated with the four point
function of the $\Phi$ field. By power counting, the four point
function may be logarithmically divergent but this divergence is
canceled.  To see how this cancellation happens, consider the highest degree 
contribution in the internal (loop) momentum. It
contains four $D^2$ factors but only two of them can be converted into
momentum since one of those
 remaining must be associated to the loop integration
to produce a nonvanishing result. Therefore, one of the $D^2$ factors does not contribute to
the degree of divergence which becomes less by one than it was
initially thought. Hence, the resulting contribution is finite.

From a formal standpoint, the  divergences  we found  may be absorbed into
field and coupling constant redefinitions

\begin{eqnarray}
\Phi_{i0}&=&Z_{\Phi}^{1/2}\Phi_i,\nonumber\\
\Sigma_0&=&Z_{\Sigma}\Sigma,\nonumber\\
g_{0}&=&Z_{g} g,
\end{eqnarray}

\noindent
where $\Phi_{i0},\Sigma_0,g_0$ are the bare quantities and $\Phi_i,\Sigma,g$ 
the renormalized ones. From (\ref{6}) we obtain

\begin{equation}
Z_{\Phi} =1+ \frac{8}{\pi^2 N \epsilon}, 
\end{equation}

\noindent
which, as mentioned earlier, is the same as in the commutative
case. Nevertheless, unlike the commutative case, the $Z_{\Sigma}$
renormalization constant is not trivial. In fact, from (\ref{9}) we
have

\begin{equation}
Z_{\Sigma}Z_{\Phi}= 1 + \frac{6}{\pi^2 N \epsilon} ,
\end{equation}

\noindent
from which

\begin{equation}
Z_{\Sigma}= 1 -\frac{2}{N\pi^2\epsilon}.
\end{equation}

\noindent
$Z_g$ is fixed by the mass gap equation. As this equation is not
modified by the effect of the noncommutativity, an identical mass formula
is to be expected.  Indeed, the condition that
$\Sigma$ has  zero vacuum expectation value leads to

\begin{equation}
\frac{i}{Z_{g}g}+ Z_\Phi\int\frac{d^3k}{(2\pi)^3}\frac{1}{k^2-m^2}=0,
\end{equation}

\noindent
which after integration  becomes

\begin{equation}
\frac{1}{g} = -Z_\Phi Z_{g} \frac m{4\pi},
\end{equation}

\noindent
fixing  $Z_\Phi Z_{g}$ to be a finite constant. By making the choice

\begin{equation}
Z_\Phi Z_{g}= 1-\frac{\mu}m,
\end{equation}

\noindent
we obtain

\begin{equation}
\frac{1}{g} =\frac{\mu}{4\pi} -\frac{m}{4\pi},
\end{equation}

\noindent
in agreement with the commutative case \cite{prev,Koures}.  From this,
it also follows that the theory is asymptotically free, the
renormalization group beta function being given by

\begin{equation}
\beta(g)= -\frac{\mu}{4\pi} g.
\end{equation}

We are now in a position to prove that the model is renormalizable at
any finite order of $1/N$. The first observation is that supergraphs with
two external $\Sigma$ lines have negative degree of divergence. Indeed, the
number of vertices $V$ in such graphs is always even, $V-1$ is odd
 and, then, the number of $D^2$ factors that may be turned
into momenta decreases by one from the value specified in
(\ref{8}). Hence, these supergraphs are superficially convergent.
By the same reason,  all supergraphs with two external $\Phi$
lines are at most logarithmically divergent. Indeed, the number of vertices
is also even leading to the conclusion that one of the $D^2$ factors is
superfluous and can not be converted into momentum. Thus, the superficial degree of divergence decreases from one to zero. Analogous reasoning applied to the
four point function of the $\Phi$ field shows that there is no
overall divergence associated to the ultraviolet behavior of the graph as 
whole. We may conclude that there are at most logarithmic divergences and, 
therefore, only a mild integrable infrared singularity will appear. This is a power counting renormalization 
condition, which  is necessary but not sufficient to guarantee that the model is 
renormalizable.  It still remains to prove that the needed counterterms
have the same Moyal product structure of those vertices already present in the
original action. Specifically,  one needs to show that, at any given order
of $1/N$,  the divergent
parts of the supergraphs with two $\Phi$ and one $\Sigma$ external  lines 
generate a counterterm of the form $\int d^5 z \Sigma\star\Phi_i\star\Phi_i$.
This result follows from the statements:

1. Loop integrations associated to nonplanar subgraphs are finite. Indeed,
any such (sub)graph contains a phase factor of the form \cite{Filk}

\begin{equation}
\label{phase}
\Gamma=V(k,p)\exp(\frac{i}{2}\sum_{i,j}I_{ij}k_i\wedge k_j),
\end{equation} 

\noindent
where $k$ and $p$ denote the sets of internal and external momenta,
$I_{ij}$ is the {\it intersection} matrix whose entries are $\pm 1$ depending 
on whether
the lines carrying momenta $k_i$ and $k_j$ cross from different sides
and $0$ otherwise. Thus, the analytic expression associated to
such nonplanar subgraph must contain a factor of the form

\begin{equation}
\int \frac{d^3 k_i}{(2\pi)^3}\frac{1}{(k^2_i-m^2)^\alpha}e^{ik_i\wedge l}
(\ldots),
\end{equation}

\noindent
where the dots indicate terms which do not depend on $k_i$, $l$ is some
linear combination of the external and internal momenta, and 
$\alpha$ is either an integer or half integer number.  From (\ref{macd})
we see that the $k_i$ integral is finite and furthermore that
the $l$ integration will contain a convergence factor as well. This shows that
nonplanar subgraphs are superficially convergent. An immediate consequence of this fact
is that the degree of divergence of an arbitrary graph is determined only
by its planar subgraphs.

2. All divergent contributions associated to graphs with $N_{\Sigma}=1$ 
and $N_\Phi=2 $ have the Moyal product structure $\int \Sigma\star\Phi\star\Phi$.
This result follows from the property

\begin{equation}\label{last}
\cos a_1 \cos a_2 \ldots \cos a_n = \frac1{2^{n-1}} \sum \cos(a_1
\pm  a_2\ldots\pm  a_n), 
\end{equation}

\noindent
where the sum is taken over all possible combinations of the $\pm$
signs. The above expression allows one to demonstrate that for any
graph with an arbitrary number of loops there is one planar
contribution, i.e., containing a cosine factor depending only on the external
momenta. In fact, this result holds for an arbitrary (having any
number of external lines) one-loop graph. To see that, consider the
one-loop graph depicted in Fig. 4. From (\ref{last}) we extract the
following term

\begin{equation}
\cos[ k\wedge p_1+(k+p_1)\wedge p_2+\ldots + (k+p_1+\ldots+p_{n-1})\wedge p_n],
\end{equation}

\noindent
which, after taking into account external momentum conservation, turns
 out not to depend on the loop momentum, as stated. We now assume that
 the result holds for an arbitrary $n$-loop graph.  We may then
 increase the number of loops by one unity by joining two external
 lines through a tree diagram consisting of one line with possibly
 other external lines attached to it (see Fig. 5). Using the same
 procedure as in the one-loop case, we can verify again that
 there is one term whose cosine factor does not depend on the new loop
 momenta.  This proves our statement.

As a concluding remark, we want to emphasize that, in spite of the
nonlocality of the Moyal interaction, the noncommutative nonlinear
sigma model remains renormalizable to all orders of $1/N$, as we
proved.  The renormalization program is nevertheless modified as
compared with the commutative case since a renormalization for the
$\Sigma$ field is now required. Furthermore, the leading order
correction, which is of order $N$, does not depend on the
noncomutativity parameter. This dependence occurs in the next to the
leading order and it is of the form $1/\sqrt {\Theta^2}$, hence it
does not possess a commutative limit.

This work was partially supported by Funda\c c\~ao de Amparo \`a
Pesquisa do Estado de S\~ao Paulo (FAPESP) and Conselho Nacional de
Desenvolvimento Cient\'\i fico e Tecnol\'ogico (CNPq). Two of us
(H.O.G. and V.O.R.) also acknowledge support from PRONEX under contract
CNPq 66.2002/1998-99. A. Yu. P. has been supported by FAPESP, project
No. 00/12671-7.

\begin{figure}[h]
\centerline{\epsfbox{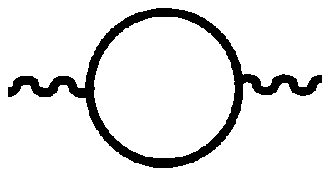}}
\caption{Leading contribution to the two point proper function of the $\Sigma$ field. Continuous and wavy lines represent the $\Phi$ propagator and the
external $\Sigma$ field.}  
\label{adilson}
\end{figure}
\begin{figure}[h]
\centerline{\epsfbox{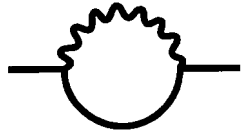}}
\caption{$1/N$ correction to the propagator of the $\Phi$ field.}
\end{figure}
\begin{figure}[h]
\centerline{\epsfbox{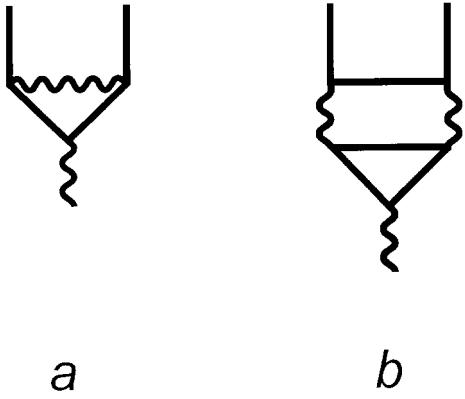}}
\caption{Leading vertex corrections.}
\end{figure}
\begin{figure}[h]
\centerline{\epsfbox{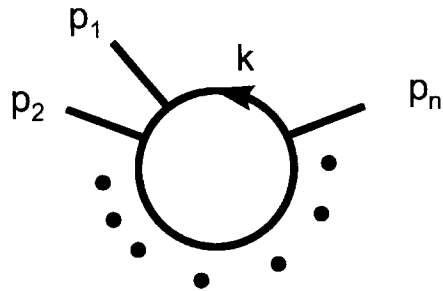}}
\caption{One-loop contribution to the  Green function.}
\end{figure}
\begin{figure}[h]
\centerline{\epsfbox{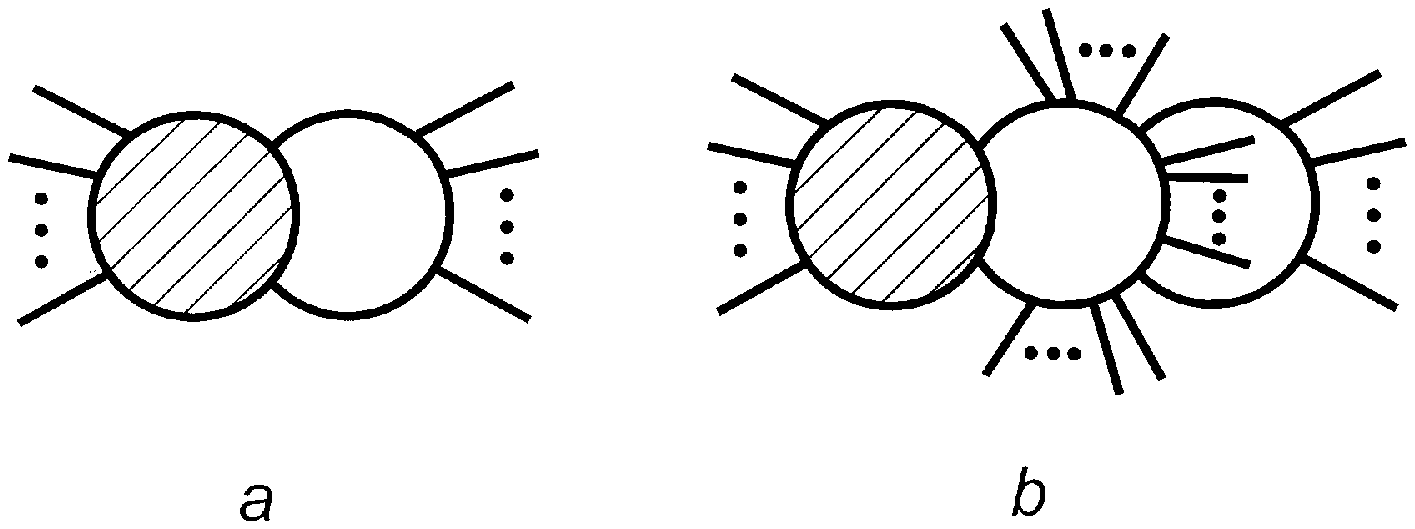}}
\caption{(a) An $n$-loop contribution. The dashed circle stands for an arbitrary
graph. (b) The $n+1$-loop graph obtained by joining two external lines through
a tree structure.}
\label{adilson-2}
\end{figure}

\begin{references}

\bibitem{Minwalla} S. Minwalla, M. V. Raamsdonk and N. Seiberg,
``Noncommutative Perturbative Dynamics'', JHEP {0002}, 020 (2000),
hep-th/9912072; M. V. Raamsdonk and N. Seiberg,``Comments on Noncommutative Perturbative  Dynamics'', JHEP {0003}, 035 (2000), hep-th/0002186.

\bibitem {Matusis} A. Matusis, L. Susskind and N. Toumbas, ``The IR/UV
Connection in the Noncommutative Gauge Theories'', JHEP {0012}, 002
(2000), hep-th/0002075.

\bibitem{Hayakawa} M. Hayakawa, ``Perturbative Analysis on Infrared 
Aspects of Noncommutative QED on $R^4$'', Phys. Lett. B478 (2000) 394,
 hep-th/9912094; ``Perturbative Analysis on Infrared and
Ultraviolet Aspects of Noncommutative QED on $R^4$'',  hep-th/9912167.

\bibitem {Arefeva2} I. Ya. Aref'eva, D. M. Belov, A. S. Koshelev and
O. A. Rytchkov, ``Renormalizability and UV/IR Mixing in Noncommutative
Theories with Scalar Fields'', Phys. Lett. B487 (2000) 357.

\bibitem{Arefeva} I. Ya. Aref'eva, D. M. Belov and A. S. Koshelev, ``UV/IR 
Mixing for Noncommutative Complex Scalar Field Theory. 2. (Interaction
with Gauge Fields'', Nucl. Phys. Suppl. 102, 17 (2001), hep-th/0003176.

\bibitem{Martin} C.~P.~Martin and D.~Sanchez-Ruiz,
``The one-loop UV Divergent Structure of U(1) Yang-Mills Theory on
Noncommutative $R^4$'' , 
Phys. Rev. Lett.  {83} (1999) 476, hep-th/9903077.

\bibitem{Grosse} H.~Grosse, T.~Krajewski and R.~Wulkenhaar,
``Renormalization of Noncommutative Yang-Mills Theories: a Simple
Example'', 
hep-th/0001182.

\bibitem{Armoni} A. Armoni, ``Comments on Perturbative Dynamics of
Noncommutative Yang-Mills Theory'', hep-th/0005208.

\bibitem{Bonora} L.~Bonora, M.~Schnabl and A.~Tomasiello, ``A Note on
Consistent Anomalies in Noncommutative YM Theories'', Phys. Lett. B485
(2000) 311, hep-th/0002210.

\bibitem{Sheikh}  M.~M.~Sheikh-Jabbari,
``Renormalizability of the Supersymmetric Yang-Mills Theories on the
Noncommutative Torus'', 
JHEP { 9906}, 015 (1999),  hep-th/9903107.

\bibitem{Chepelev} I.~Chepelev and R.~Roiban,
``Renormalization of Quantum Field Theories on Noncommutative $R^d$.
I: Scalars'', JHEP 0005 037 (2000), hep-th/9911098; ``Convergence Theorem
for Noncommutative Feynman Graphs and Renormalization'', JHEP 0103 001
(2001). 

\bibitem{Gracia} J.~M.~Gracia-Bondia and C.~P.~Martin, ``Chiral Gauge
Anomalies on Noncommutative $R^4$,'' Phys. Lett. B479 (2000) 321,
hep-th/0002171.

\bibitem{Douglas} For a review see M. R. Douglas and N. A. Nekrasov, ``Noncommutative Field Theory'', hep-th/0106048, to appear in Rev. Mod. Phys.
 
\bibitem{Girotti} H. O. Girotti, M. Gomes, V. O. Rivelles and A. J. da Silva,
``A Consistent Noncommutative Field Theory: The Wess--Zumino Model
Nucl. Phys. B587, 299 (2000), hep-th/0007080.

\bibitem{prev} H. O. Girotti, M. Gomes, V. O. Rivelles and  A. J. da Silva,
``The Noncommutative  Supersymmetric Nonlinear Sigma  Model'', hep-th/0102101.


\bibitem{Popp} A. Bichl, J. M. Grimstrup, H. Grosse, L. Popp, M. Schweda
and R. Wulkenhaar, ``The Superfield Formalism Applied to the Noncommutative
Wess--Zumino Model'', JHEP 0010: 046, 2000;  hep-th/0007050.

\bibitem{Zanon} D. Zanon,''Noncommutative $N=1$, $N=2$ Super $U(N)$ Yang--Mills: UV/IR Mixing and Effective Action Results at One Loop'', Phys Lett. B502, 265 (2001); ``Noncommmutative Perturbation in Superspace'', Phys. Lett. B504, 101
(2001).


\bibitem{ours} I.L. Buchbinder, M. Gomes, A.Yu. Petrov and V.O Rivelles,
``Superfield Effective Action in the Noncommutative Wess-Zumino Model'', 
hep-th/0107022, to appear in Phys. Lett. B.


\bibitem{Koures} V.G. Koures and  K.T. Mahantappa, Phys. Rev. D43, 3428
(1991); D45, 480 (1992).


\bibitem{Gelfand} I. M. Gel'fand and G. E. Shilov, ``Generalized Functions'',
Vol. 1 (Academic Press, 1964).

\bibitem{Grisaru} S. J. Gates Jr., M. T. Grisaru, M. Rocek and W. Siegel,
``Superspace or One Thousand and One Lessons in Supersymmetry'' (The 
Benjamin/Cummings Pub. Co, 1983).

\bibitem{Filk} T. Filk, Phys. Lett. B376, 53 (1996).
\end{references}
\end{document}